\documentclass[english,aps,prl,onecolumn, superscriptaddress,showpacs, amsmath]{revtex4-1}

\usepackage[T1]{fontenc}
\usepackage[latin9]{inputenc}
\usepackage{color}
\usepackage{units}
\usepackage{amssymb}
\usepackage{graphicx}
\usepackage{esint}
\usepackage{bm}
\usepackage{natbib}
\usepackage{amsmath}
\usepackage[colorlinks=true, citecolor=blue]{hyperref}
\usepackage{lipsum, babel}
\setcitestyle{journalcolor= blue}

\usepackage{babel}

\begin{document}
\title{Andreev reflection near the Dirac point at Graphene - NbSe$_2$ junction}

\author{Manas Ranjan Sahu}
\affiliation{Department of Physics, Indian Institute of Science, Bangalore 560012, India}
\author{Pratap Raychaudhuri}
\affiliation{Tata Institute of Fundamental Research, Homi Bhabha Road, Colaba, Mumbai 400 005, India}
\author{Anindya Das}
\email{anindya@physics.iisc.ernet.in}
\affiliation{Department of Physics, Indian Institute of Science, Bangalore 560012, India}

\begin{abstract}
Despite extensive search for about a decade, specular Andreev reflection is only recently realized in bilayer graphene-superconductor interface. However, the evolution from the typical retro type Andreev reflection to the unique specular Andreev reflection in single layer graphene has not yet been observed. We investigate this transition by measuring the differential conductance at the van der Walls interface of single layer graphene and $NbSe_2$ superconductor. We find that the normalized conductance (G$_{T<T_c}$/G$_{T>Tc}$) becomes suppressed as we pass through the Dirac cone via tuning the Fermi level and bias energy, which manifests the transition from retro to non-retro type Andreev reflection. The suppression indicates the blockage of Andreev reflection beyond a critical angle ($\theta_{c}$) of the incident electron with respect to the normal between the single layer graphene and the superconductor junction. The results are compared with a theoretical model of the corresponding setup.
\end{abstract}
\maketitle
Andreev reflection (AR)~\cite{andreevReflection1964} is the underlying quantum phenomena by which the current flows from a normal (N) region into a superconductor (S) at the normal-superconductor junction. In this process the dissipative current from the normal side converts into dissipation-less super-current in the superconductor. Microscopically an incident electron from the normal side makes a pair with another electron below the Fermi energy ($E_F$) to form Cooper pair at N-S junction. As a result a hole reflects back to the normal side by retracing back the path of the incident electron, which is known as retro Andreev reflection (RAR). However, there can be a physical phenomena in which the conversion of incident electron into the reflected hole is not in the same path, which is known as specular Andreev reflection (SAR) at N-S junction. SAR was not predicated until Beenakker ~\cite{PhysRevLett.97.067007,beenakker2008colloquium} discovered that this rare phenomena is possible for relativistic electrons in graphene with a superconducting interface. Since then there have been many proposals about SAR in different systems like topological insulator-superconductor junction~\cite{TIAndreevReflection} and two-dimensional semiconductor with spin-orbit coupling and d-wave superconductor~\cite{SOCAndreevReflection} etc. However, experimentally it remains challenging to observe this effect.

It was pointed out in Ref ~\cite{PhysRevLett.97.067007,beenakker2008colloquium} that in case of single layer graphene (SLG), when the chemical potential ($\mu$) of the superconductor is far away from the Dirac point ($E_F$ of SLG >> $\Delta$) the AR will be retro type, which is an intra-band process as shown in Fig. 1a. There will be a critical value ($\theta_{c}$) of the incident electron at the N-S junction beyond which the AR will be blocked. The $\theta_{c}$ emerges as a consequence of the conservation of momentum between the incident electron above $E_F$ and reflected hole below $E_F$. When the $E_F$ of SLG >> $\Delta$ the $\theta_{c}$ tends to $\pi$/2, which means the electron can reflects back as a hole with any incident angle from zero to $\pi$/2. The value of the $\theta_{c}$ will decrease as the $E_F$ approaches towards the Dirac point. As a result AR will be progressively suppressed while passing through the Dirac point and AR will be no longer RAR because the path of incident electron and reflected hole will be different. This is the onset of specular type AR. However, when the Dirac point of SLG is within the  superconducting gap and $E_F$ < $\Delta$, the $\theta_{c}$ will again tend to $\pi$/2 and as a result AR will be enhanced. In the latter one the AR (Fig. 1a) is inter-band process in which there is a sign change of the hole mass and will be true SAR.

\begin{figure*}[ht!]
\includegraphics[width=1\textwidth]{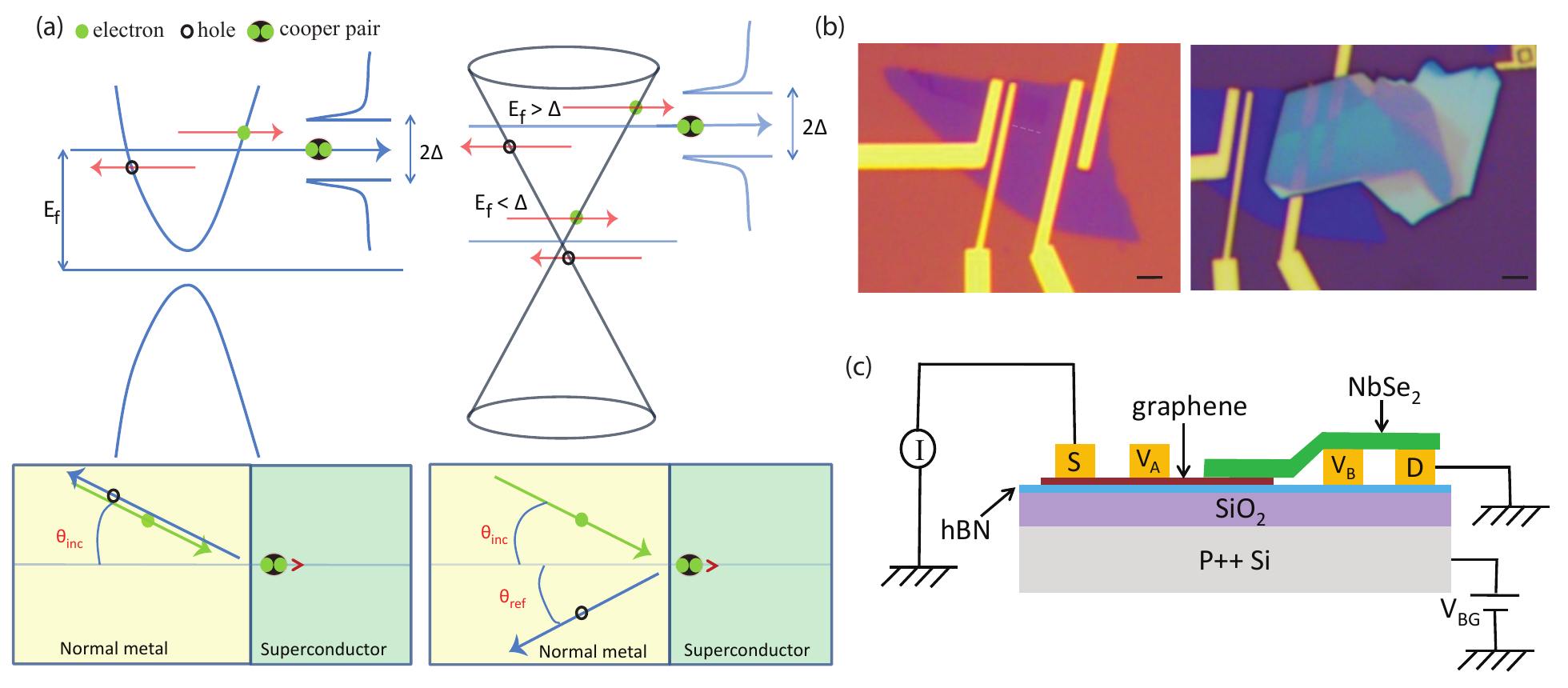}
 \caption{(Color Online) (a) (top left) AR for a semiconductor-superconductor junction, (bottom left) schematic of corresponding retro reflection process. (top right) In case of graphene another type of AR  appears when the Fermi energy is very close to Dirac point, (bottom right) schematic of corresponding specular AR process. (b) (left) Optical image of graphene on $hBN$ and (right) the device with $NbSe_2$ Scale bar 2$\mu m$. (c) Schematic of the measurement setup.}
 \label{fig1}
\end{figure*}

\begin{figure*}[ht!]
 \includegraphics[width=1\textwidth]{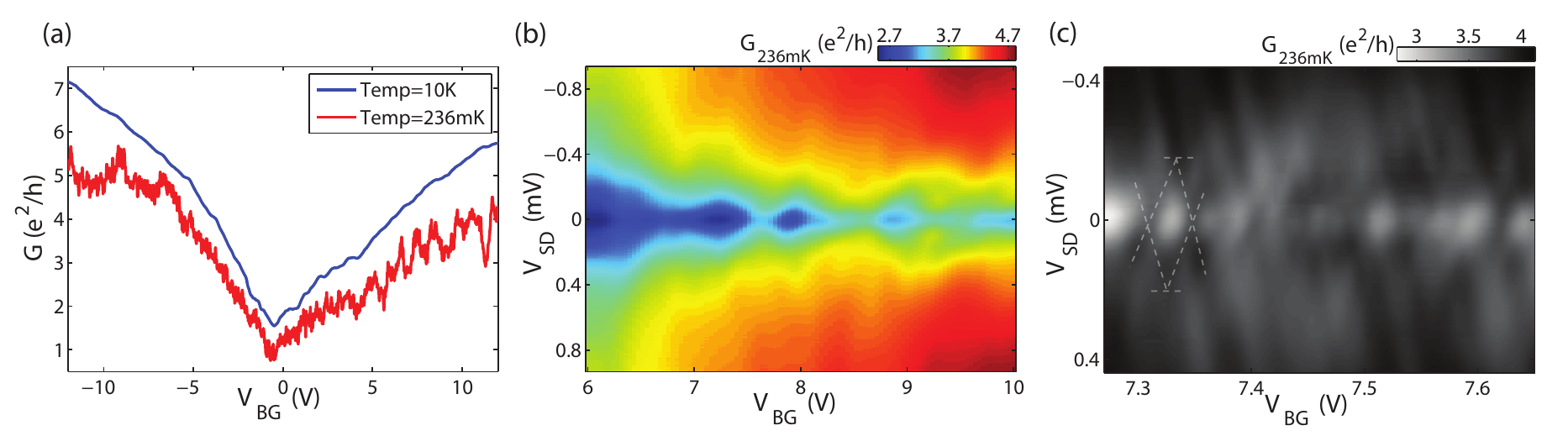}
 \caption{(Color Online) (a) Gate responses at 10K and 236mK. (b) 2D colormap of $dI/dV$ as a function of V$_{BG}$ and V$_{SD}$ showing Fabry Perot oscillations. (c) High resolution plot of $dI/dV$ with diamond structures. The dashed white lines indicate the periodicity and the energy scale of the F.P. oscillations.}
 \label{fig:example}
\end{figure*}

The main difficulties to observe SAR experimentally is the Fermi energy broadening ($\delta E_F$). For example graphene on $Si/SiO_2$ substrate has $\delta E_F$ of the order of $\sim$ 100 meV due to charge puddles present on the substrate. Very recently, hexagonal boron nitride ($hBN$) supported bilayer graphene (BLG) has been used to reduce the $\delta E_F$ and the signature of SAR has been observed experimentally ~\cite{efetov2015specular} after a decade of its theoretical prediction~\cite{PhysRevLett.97.067007,beenakker2008colloquium}. However, for bilayer graphene the energy dispersion is quadratic near the Dirac point as well as there could be band gap opening at charge neutrality point. On the other hand the dispersion in SLG is linear having relativistic massless Dirac fermion characteristics, which gives rise to Klein tunneling at the interface of a p-n junction. Because of this property, even if there is a work function mismatch between the SLG and SLG underneath of superconductor, there will be electron transport through the interface by virtue of AR. Therefore, investigating the transition from RAR to SAR in SLG would help to understand the underlying Andreev process in graphene-superconductor junctions. In this article we have carried out the transport measurements on normal - SLG - superconductor junction, where the SLG has been supported by $hBN$ and achieve $\delta E_F$ $\sim$ 10 meV. As a superconductor, $NbSe_2$ has been used whose superconducting gap 2$\Delta$ $\sim$ 2 meV. The conductance ($G$ = $dI/dV$) measurements with the carrier density and bias reveal that the normalized conductance ($G_{T<T_c}/G_{T>T_c}$) becomes suppressed as we pass through the Dirac point. The suppression of conductance around the Dirac point matches fairly well with our theoretical calculation based on Blonder-Tinkham-Klapwijk (BTK) formalism~\cite{blonder1982transition,beenakker2008colloquium}. Our results with the theoretical support unveil the onset of transition from retro AR to non retro type AR in SLG-superconductor interface.

Fig. 1b shows an optical image of a SLG on $hBN$. A thin layer of hBN ($\sim$ 10nm) was first exfoliated on a $Si/SiO_2$ wafer. This was followed by transfer of SLG on $hBN$ by dry transfer technique~\cite{dean2010boron}. The contacts for SLG and predefined contacts for $NbSe_2$ were made of Cr/Au(5/70 nm) using standard electron beam lithography. At the last step a thin $NbSe_2$ ($\sim$ 20nm) was transferred on SLG and predefined contacts, as shown in Fig. 1b. 
We should mention that in order to avoid the oxidization of the bottom surface of $NbSe_2$, the $NbSe_2$ was transferred within few minutes after fresh exfoliation of a $NbSe_2$ bulk flake. The schematic of the measurement is shown in Fig. 1c, where the conductance between the normal(Au) - SLG - superconductor ($NbSe_2$) has been measured using conventional lock-in technique. All the measurements were carried out in a $^3He$ refrigerator having base temperature of 236mK. The typical contact resistances between Au-graphene and Au-$NbSe_2$ are less than a hundred of Ohm where as the graphene - $NbSe_2$ interfaces show $\sim$ 1.5-3.0 kOhm contact resistance (details in supplementary information - Fig. S1). In this article, we have repeated the experiments for two more representative devices. In the S.I.(Fig. S2), we have also shown the characterization of $NbSe_2$ thin flake transferred on predefined gold contacts and found the $2\Delta$ $\sim$ 2meV, which also match very well with the critical temperature (T$_c$ $\sim$ 6.5K) measurement.

\begin{figure*}[ht!]
\includegraphics[width=1\textwidth]{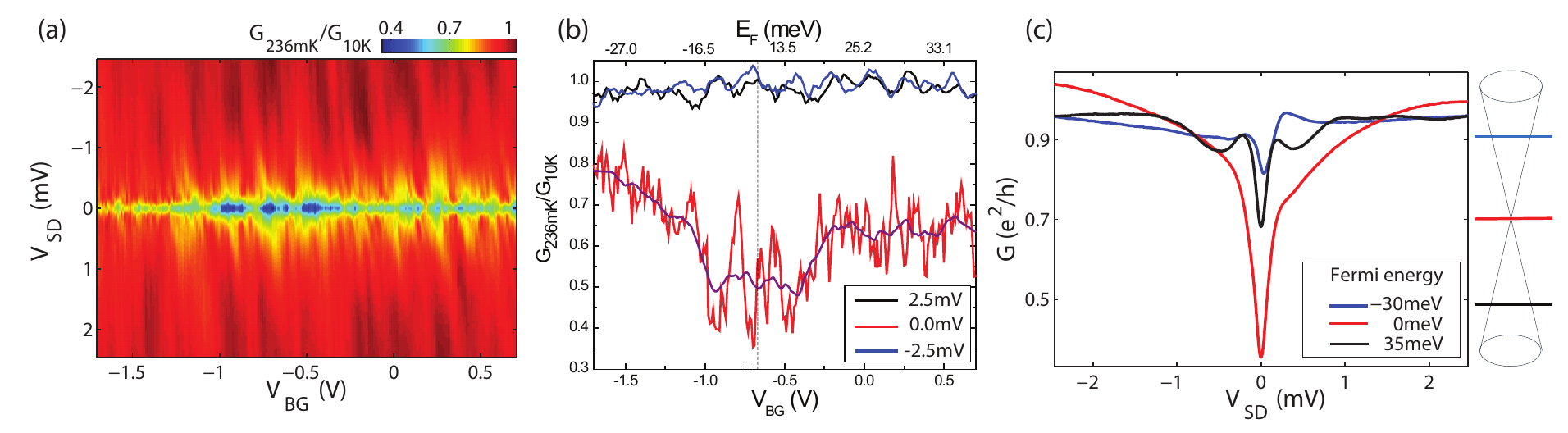}
 \caption{(Color Online) (a) 2D colormap of normalized differential conductance as a function of V$_{BG}$ and V$_{SD}$. (b) G$_{236mK}$/G$_{10K}$ versus V$_{BG}$ plots for different values of V$_{SD}$. The brown solid line corresponds to average over 20 raw data points (red line) at V$_{SD}$ = 0 mV. The vertical dashed line indicates the Dirac point. (c) G$_{236mK}$/G$_{10K}$ versus V$_{SD}$ plots for different values of V$_{BG}$ at -1.65V, -0.65V and 0.65V, which correspond to $E_F$ at -30 meV, 0 meV and 35 meV, respectively. The horizontal solid lines schematically indicate those energies in a Dirac cone.}
 \label{fig:example}
\end{figure*}

Fig. 2a shows the $dI/dV$ of one of our representative devices as a function of gate voltage ($V_{BG}$) at 10K and 236mK. The conductance almost increases linearly with $V_{BG}$. By analyzing the 10K data we extract the mobility ($\mu$), in-homogeneity ($\delta n$) and $\delta E_F$ of the device, which are 12,000 cm$^2$V$^{-1}$s$^{-1}$, 10$^{10}$cm$^{-2}$ and 10meV, respectively (for details Fig. S3). The most distinct differences at two temperatures are the fine oscillations at 236mK. In order to understand the origin of these quasi periodic oscillations we have carried out the $dI/dV$ as a function of $V_{BG}$ and $V_{SD}$ (bias), shown as 2D colormaps in Fig. 2b and Fig. 2c. Fig. 2c clearly shows diamond type of oscillation, which are signature of Fabry-Perot (F.P.) oscillations and reported in graphene Josephson junctions by many groups ~\cite{rocha2010ac,allen2015visualization,shalom2015quantum,BallisticJosephson}. The F.P oscillation is due to the formation of standing waves between the normal - SLG and SLG - superconductor interface, which fairly match with our device dimension (for details Fig. S4). The above results show the ballistic nature of our normal-SLG-superconductor device at low temperature. 

It can be seen from Fig. 2b and 2c that at 236mK the $dI/dV$ has always lower value at $V_{SD}$ = 0 compared to finite $V_{SD}$ irrespective of $V_{BG}$. For T > T$_c$, $dI/dV$ is almost independent of $V_{SD}$ (S.I.). With lowering the temperature, the dip at $V_{SD}$ = 0 starts appearing below T$_c$ (S.I.-Fig. S3). At 236mK the superconducting gap at $V_{SD}$ $\sim$ $\pm$ 1meV with the subgap features at $V_{SD}$ $\sim$ $\pm$ 0.3meV are observed (S.I.). As discussed before the subgap features are due to F.P. oscillations. Within the superconducting gap the electron can transport by Andreev reflection. For a transparent (barrier strength, Z=0) SLG-superconductor interface the conductance should double because of equal contribution coming from the reflected hole, which adds up as excess current. However, for an interface with Z > 0.5 there will be a decrease in conductance within the superconducting gap. Our data suggests Z $\sim$ 0.7, which corresponds to a transparency, T $\sim$ 0.7 (T = 1/1+Z$^2$) at SLG-superconductor interface~\cite{blonder1982transition} (S.I for details). In order to see the crossover from RAR to SAR around the Dirac point we plot the normalized 2D colomap ($G_{236mK}/G_{10K}$) in Fig. 3a. It can be clearly seen that above superconducting gap the normalized conductance is independent of $V_{BG}$ but close to zero bias ($V_{SD}$ $\leq$ 0.3mV) the normalized conductance is suppressed around the Dirac point. This is clearly visible in Fig. 3b, where the normalized conductance is plotted as a function of $V_{BG}$ for $V_{SD}$ = 0 and $V_{SD}$ = $\pm$ 2.5mV. The similar suppression can be also seen if we take the vertical cuts from Fig. 3a at different $V_{BG}$ as shown in Fig. 3c, where the $G_{236mK}/G_{10K}$ at $V_{SD}$ = 0 is minimum close to the Dirac point.

\begin{figure*}[ht!]
\includegraphics[width=1\textwidth]{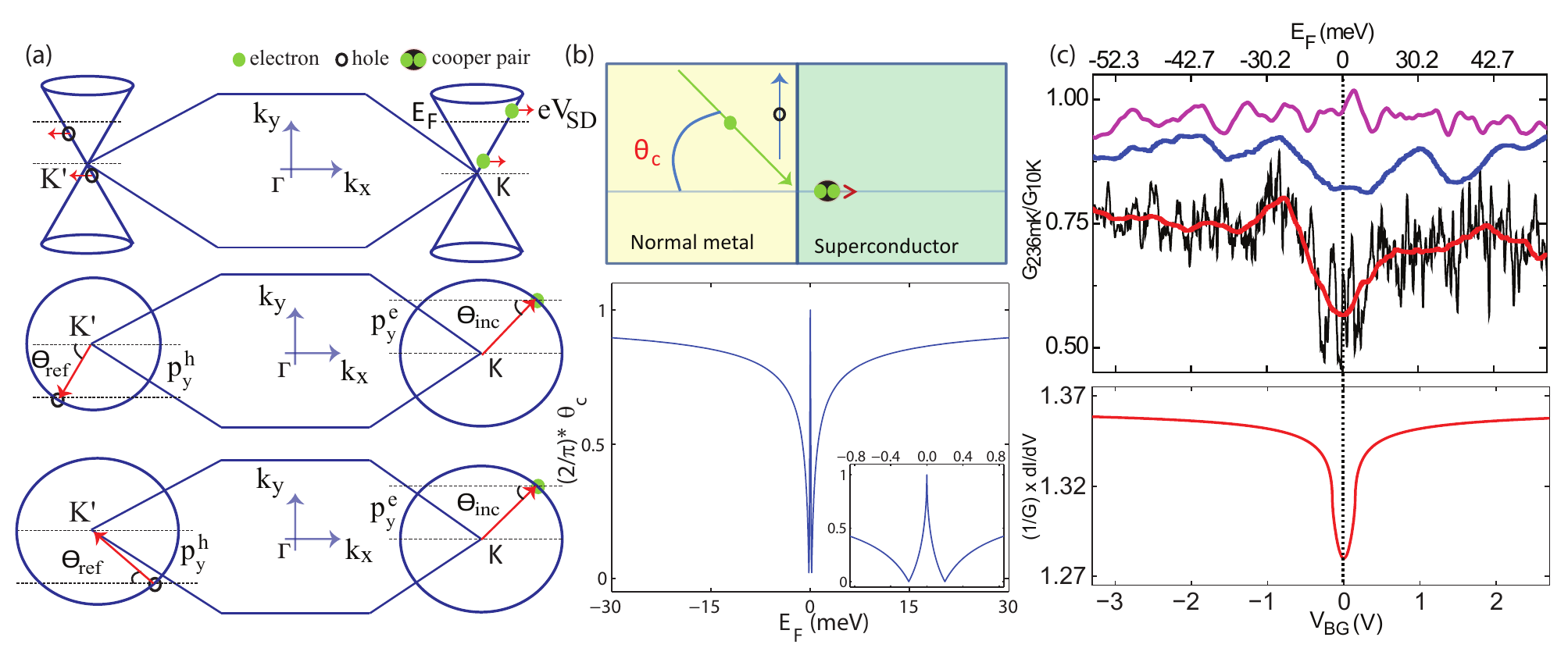}
 \caption{(Color Online) (a) (top) Schematic of AR for an electron having different Fermi energy positions. Here we have used only electronic picture i.e the Andreev reflected hole is represented as a vacancy in the electronic band at an appropriate position. The position of the incident electron and Andreev reflected hole in K$_x$ - K$_y$ plane for RAR (middle) and SAR (bottom). In all these representations arrows indicate the real-space velocity of the quasi particles. (b) (top) Schematic of the RAR process when $\theta_{inc}$ = $\theta_{c}$. The normalized value of critical angle as a function of $E_{F}$ for V$_{SD}$ = 0.2meV (bottom). Inset shows the zoomed one around the Dirac point. (c) (top) The experimental G$_{236mK}$/G$_{10K}$ data (black line) as a function of V$_{BG}$, similar to Fig. 3b. The red solid line corresponds to the average over 200 raw data points. The blue solid line (shifted vertically by 0.15) is the experimental data of another SLG-NbSe$_2$ device with higher $\delta E_F$ $\sim$ 30 meV. The magenta solid line (shifted vertically by 0.2) corresponds to G$_{236mK}$/G$_{10K}$ of SLG alone and there is no suppression around the Dirac point. Theoretically calculated normalized differential conductance of SLG-NbSe$_2$ as a function of V$_{BG}$ with $\delta E_F$ $\sim$ 12 meV for V$_{SD}$ = 0.2meV (bottom). The vertical dashed line indicates the position of Dirac point.}
 \label{fig:example}
\end{figure*}

For AR at SLG-superconductor interface, an incident electron above the Fermi energy ($E_{F}+eV_{SD}$) makes a Cooper pair with an electron below the Fermi energy ($E_{F}-eV_{SD}$) having opposite spin and opposite valley~\cite{PhysRevLett.97.067007}, as shown in Fig. 4a (top). The missing electron below $E_{F}$ returns as hole and thus the current flow at the interface. The component of momentum ($P_{y}$) along the interface must be strictly conserved and the component of hole velocity perpendicular to the interface must be negative to that of incident electron. These two conditions will restrict the hole to be at a specific point in $K_x$-$K_y$ plane, as shown in fig 4(a) (middle). As the radius of constant energy contour decreases with decreasing Fermi energy, the angle of reflected hole ($\theta_{ref}$) will be larger than the angle of incident electron ($\theta_{inc}$). Therefore, beyond a critical angle ($\theta_{c}$) of incident electron the $\theta_{ref}$ will exceed 90 degree and thus AR process will stop (Fig. 4b - top). The $\theta_{c}$ can be expressed as following~\cite{PhysRevLett.97.067007,beenakker2008colloquium,bhattacharjee2006tunneling}.

\begin{equation}
 \Theta_c=sin^{-1} \frac{|E_F-eV_{SD}|}{E_F+eV_{SD}}
\end{equation}

Using Eqn.1 the $\theta_{c}$ (normalized by total angle $\pi$/2) as a function of $E_{F}$ is shown in Fig. 4b for $V_{SD}$ = 0.2mV. It can be seen that far away from the Dirac point the $\theta_{c}$ is close to $\pi$/2, which means the incident electron can reflect back as a hole with any incident angles ($\theta_{inc}$ from zero to $\pi$/2). With lowering the $E_{F}$ the $\theta_{c}$ decreases. However, when the $E_{F}$ is close to the Dirac point the AR is specular type (Fig. 4a - bottom) and in that case $\theta_{c}$ again becomes $\pi$/2. In order to compare with the experimental data one needs to calculate the differential conductance. For that we need to know the coefficient of andreev reflection($r_A$) as well as normal reflection($r$). These probabilities depend on the position of Fermi energy as well as on angle of incident electron ($\alpha$ = $\theta_{inc}$). The differential conductance has been numerically calculated using the following expression based on BTK formula~\cite{blonder1982transition,PhysRevLett.97.067007}.

\begin{equation}
dI/dV=G(E_F)[\int_{0}^{\theta_c}(1-|r|^2+|r_A|^2)  cos \alpha  d\alpha + \int_{\theta_c}^{\pi/2}(1-|r|^2)  cos \alpha  d\alpha ]
\end{equation}

$G(E_F)=\frac{4e^2}{h} N(E_F)$ is the ballistic conductance of graphene channel, where $N(E_F)$ is the number of transverse modes present in the graphene channel. The expression of $r_A$ and $r$ are given in the S.I.~\cite{PhysRevLett.97.067007}. Using those expressions and Eqn. 2 we have calculated $1/G(E_F)$ x $dI/dV$ as a function of V$_{BG}$ for different Fermi energy broadening, as shown in Fig. S5. To compare our experimental data (red line Fig. 4c-top) we show the theoretical calculation for $\delta E_F$ $\sim$ 12 meV in Fig. 4c - bottom. It can be seen from Fig. 4c that the experimental data qualitatively agrees with the theory. The discrepancy between the experimental and theoretical values could be due the finite barrier (Z $\sim$ 0.7) at SLG-superconductor interface, which was not considered in theory.

Now we will consider the possible effect due to the finite contact resistance (1.5-3.0 kOhm) at the SLG-superconductor interface. It is known that for graphene the contact resistance changes with $E_{F}$. However, with $E_{F}$ shift, if the contact resistance 
below and above T$_c$ changes in a similar manner (in percentage), it will not contribute to the normalized conductance ($G_{236mK}/G_{10K}$). This is justified in Fig. 3b for higher bias data. Even if there are changes in contact resistance (below and above T$_c$) with shifting $E_{F}$, it will be a gradual effect contrary to our observation in Fig. 3b for the zero bias data. We have also seen that the suppression is much weaker for SLG-NbSe$_2$ device having higher $\delta E_F$ $\sim$ 30meV (blue line in Fig. 4c-top).

In conclusion we have carried out the quantum transport measurement at SLG-NbSe$_2$ junction. Our device showing Fabry-Perot type oscillations at low-temperatures indicates the ballistic nature of our normal - SLG - superconductor device. The normalized conductance (G$_{236mK}$/G$_{10K}$) above the superconducting gap (V$_{SD}$ > $\Delta$) does not depend on the position of Fermi energy. On the other hand inside the superconducting gap the normalized conductance gets suppressed as we pass through the Dirac point. The suppression is understood in terms of blockage of AR beyond a critical angle at SLG - superconductor interface, which is also indication of 
non retro type AR because the paths for incident electrons and reflected holes are different, which is indeed the onset for SAR. However, we do not observe the signature of true specular AR at the Dirac point due to $\delta E_F$ > $\Delta$. Our experimental data matches fairly well with our theoretical calculation based on BTK formula, which will help to understand the future experiments related to this field. 

Authors thank Vivas Bagwe for preparing the NbSe2 single crystals. Authors thank Dr. Abhiram Soori, Dr. Bhaskar Kaviraj and Dr. Tanmoy Das for stimulating discussions. AD also thanks nanomission under Department of Science and Technology, Government of India for financial support.

\vskip 0.6in

\begin{center}
\textbf{SUPPLEMENTARY INFORMATION}
\end{center}

\section{Determination of contact resistance at SLG-NbSe$_2$ junction}
By measuring the resistances of our device in four probe and three probe configuration we extract out the contact resistances between SLG-gold contact and NbSe$_2$-gold contact, which are $\sim$ 100$\Omega$ and $\sim$ 70$\Omega$, respectively. In order to know the contact resistance between the SLG and NbSe$_2$ we have used the following method. We first extract out the resistivity of SLG alone from the R Vs V$_{BG}$ plot and then extrapolate for the SLG channel part between gold and NbSe$_2$. At the end the contact resistance at the SLG-NbSe$_2$ junction is determined by subtracting the SLG channel resistance from four probe gold-SLG-NbSe$_2$ resistance, which has been shown in Fig. 1a as a function of gate voltage. It can be seen that for hole and electron sides the junction resistances are $1.5 K\Omega$ and $3 K\Omega$, respectively. These values can be further justified by looking at the values of quantum Hall plateaus in SLG. Fig. 4b shows the quantum hall plateau of SLG at B = 4 Tesla. From the deviation of quantum conductance values we have evaluated the values of contact resistances. Contact resistance in the hole side is order of $1.5 K\Omega$ and in the electron side order of $3 K\Omega$, which matches fairly well with the Fig. 1a.

\begin{figure*}[ht!]
\includegraphics[width=12.4cm, height=5cm]{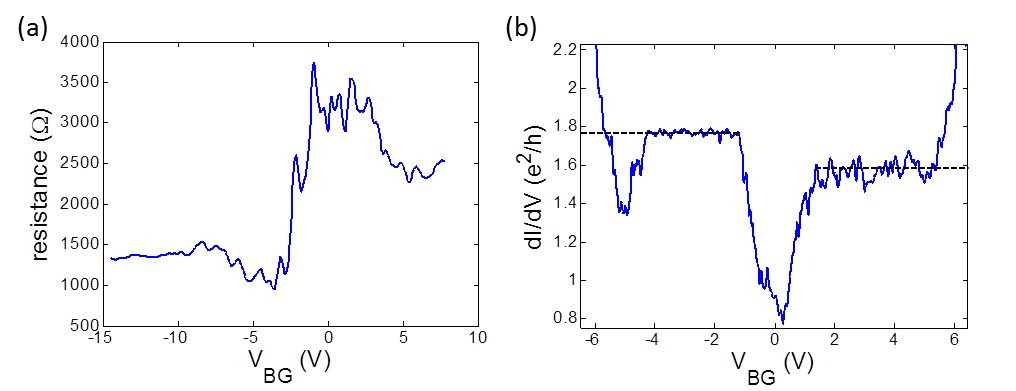}
 \caption{(Color Online) (a) SLG-NbSe$_2$ junction resistance as a function of back gate voltage (b) Quantum Hall plateau at 4 Tesla magnetic field, here the reduction from the 2e$^2$/h plateau is due to the contact resistances.}
 \label{fig1}
\end{figure*}

\section{Characterization of NbSe$_2$}
Few layer NbSe$_2$ is characterized to see the superconducting transition and to get the value of superconducting gap. One of such device is shown in Fig 2a, first predefined contacts are made and NbSe$_2$ flake is transferred at the last stage. Usually the transfer has to be done quickly, we observed that the time taken to transfer from the time of exfoliation decides the quality of contact in the devices. Our best devices in which transfer is done within 5 minutes shows contact resistance of $\sim$ 100 ohm. The four probe resistance in the device shown in Fig 2a is found to be 2.5 ohm at room temperature. We found the transition temperature around 7K, in fig 2b the resistance is plotted with time while inserting the dipstick inside the He4 Dewar. In another device we allowed the NbSe$_2$ to oxidize to form an insulating layer. In this device we did the differential conductance and found the BCS peaks at $\pm$1mV as shown in fig 2c. We observed the evolution of these peaks with temperature and found that it is vanishing above 6K.

\begin{figure*}[ht!]
\includegraphics[width=12.4cm, height=10cm]{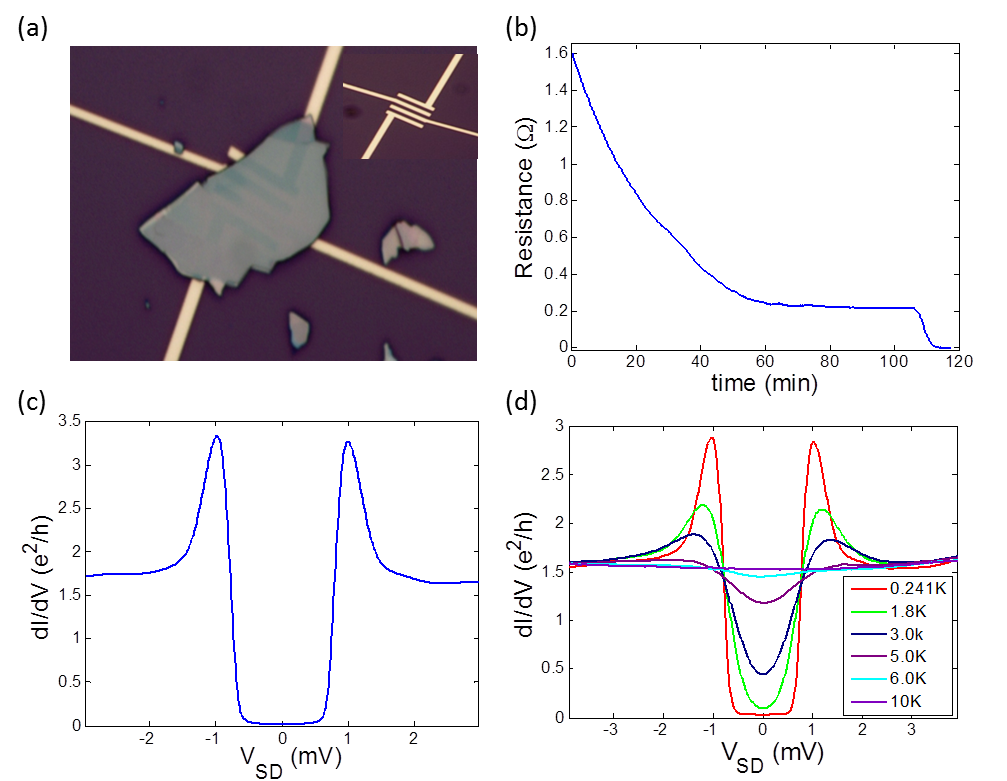}
 \caption{(Color Online) (a) $NbSe_2$ sample used for characterization which is exfoliated on predefined gold contacts as shown in the inset. (b) Resistance versus time measurement showing a sharp transition at corresponding temperature~7K. (c)Differential conductance as a function of DC bias in another sample with one tunnel barrier showing BCS peaks at $\pm$1mV. (d) Evolution of dI/dV vs DC bias plot with temperature critical temperature around 6K.}
 \label{fig1}
\end{figure*}

\section{Charge in-homogeneity and Fermi energy broadening in the sample}
To see many interesting properties of graphene it has to be suspended or supported over hBN         
~\cite{dean2010boron}. This is because when graphene is directly exfoliated on Si/SiO$_2$ substrate it becomes unevenly doped all over because of charge puddles present on the wafer which is mainly because of dangling bonds present on surface of amorphous SiO$_2$ as well as trap charges in oxides. hBN being an inert crystal improves the quality of graphene channel when supported over hBN. Fig. 3a shows the evaluation of charge in-homogeneity of the device presented in the manuscript. We have achieved $\Delta$n$_0$=10$^{10}$ cm$^{-2}$ which corresponds to Fermi energy broadening of $\pm$ 10 meV.
\begin{figure*}[ht!]
\includegraphics[width=14cm, height=5cm]{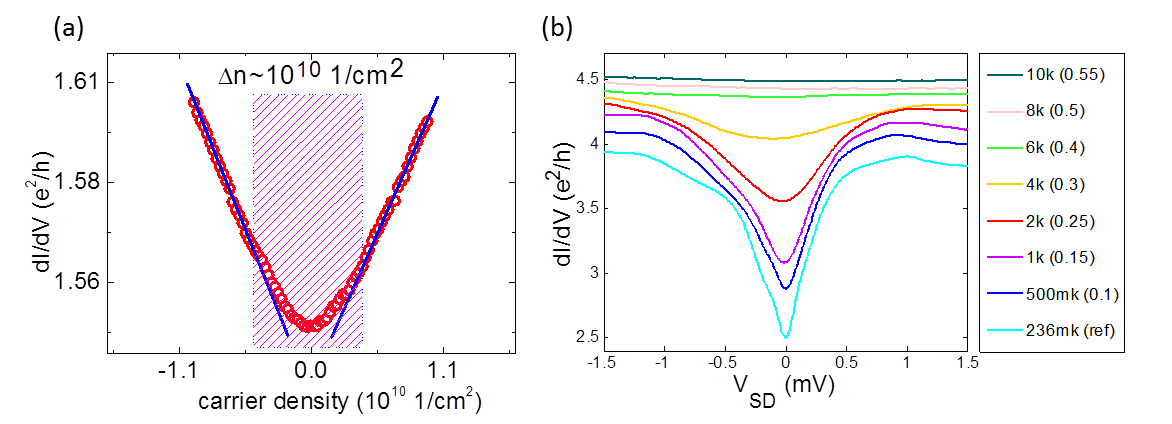}
 \caption{(Color Online) (a) Evaluation of charge in-homogenity of the device presented in the manuscript. (c) Evolution of dI/dV vs DC bias with temperature showing appearing superconductivity below 6K for SLG-$NbSe_2$ device. }
 \label{fig:example}
\end{figure*}

\section{Fabry Perot oscillations}
Due to presence of barriers at both sides of the graphene channel, it works as a Fabry Perot cavity for electrons giving rise to quasi peridic oscillations in conductance. For a specific gate voltage $V_{BG}$, carrier density$(n)=C \Delta V_{BG} /e$ where $C$ is the capacitance per unit area. Fermi wave vector$(k_F)=\sqrt{\pi n}=\frac{2\pi}{\lambda_F}$ where $\lambda_F$ is the Fermi wavelength. Condition for constructive interference is $m\lambda_F=2d$, where m is the integer and d is the channel length. This will determine the peaks in conductance plot. By knowing the position of any two peaks we can evaluate the approximate length scale the oscillations corresponds to. The major oscillations appear with periodicity of $\delta V_{BG}$ in Fig. 2b of the manuscript. $\delta V_{BG}$ corresponds to a length of $\sim$ 0.65 $\mu m$, which is close to the channel length (1.0 $\mu m$). The discrepancy has been reported earlier~\cite{allen2015visualization} and attributed to the formation of p-n junction at the interfaces as well as electric field screening, which reduces the effective length scale.

\begin{figure*}[ht!]
 \includegraphics[width=18.4cm, height=5cm]{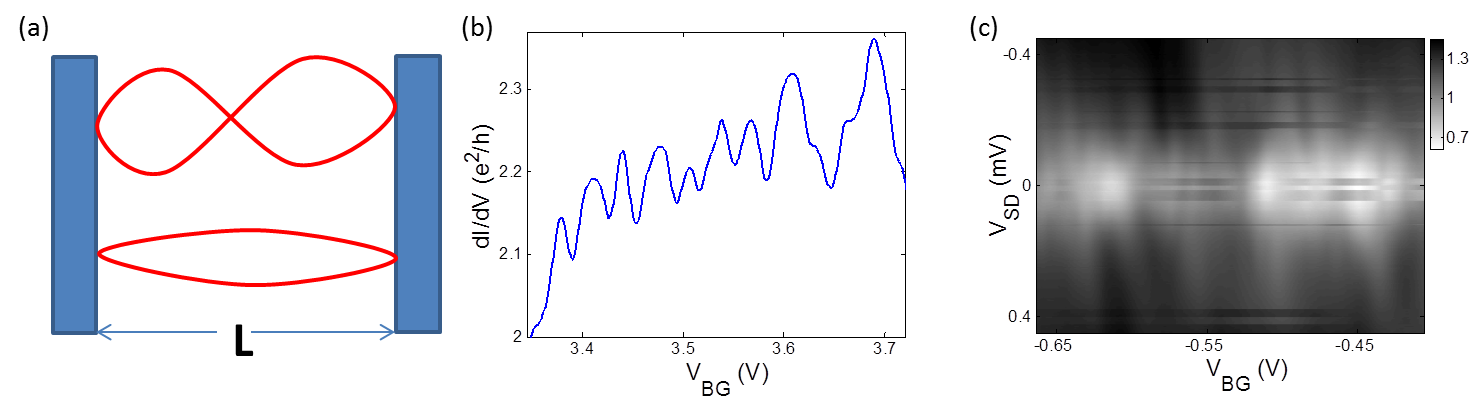}
 \caption{(Color Online) (a) Schematic of a Fabry Perot like cavity formed by barriers. (b) Fine oscillations in the gate response which evolves into Fabry Perot diamonds. (c) Fabry Perot diamonds near the Dirac point. }
 \label{fig:example}
\end{figure*}

\section{Analysis of the Transmission probability at SLG-NBSe2 junction}
The NS junction is modeled using BTK theory and the transmission probability is evaluated~\cite{blonder1982transition}. In our previously made graphene devices we found the contact resistance between gold and graphene to be hundred ohms. In this device the resistance of graphene NbSe$_2$ junction must be few kilo ohms, so we are neglecting the effect of barrier between gold and graphene. According to BKT theory normal and superconducting state transmission probabilities are given by 1/1+Z${^2}$ and 1/$(1+2Z^2)^2$ respectively. From the ratio of superconducting state conductance and normal state conductance we evaluate the value of Z=0.73. Finally we got the transmission probability to be $\sim$ 0.7.

\section{Calculation of differential conductivity}
The differential conductance through a graphene-superconductor junction can be calculated only after knowing the probabilities of Andreev reflection and normal reflection. We need to solve the Bogoliubov-De Gennes equations in both sides of the junction with appropriate boundary conditions to get the expressions of these probabilities~\cite{PhysRevLett.97.067007,beenakker2008colloquium}. The probabilities depends on the angle of incidence and the Fermi energy both. Calculation will also give a critical angle beyond which andreev reflection is not allowed as we have explained in the main text by direct observation of the electron hole conversion process. The amplitudes of Andreev reflection($r_A$) and normal reflection($r$) can be writtten~\cite{PhysRevLett.97.067007} in terms of angle of incidence($\alpha$), angle of reflection($\alpha^{\prime}$), critical angle($\alpha_c$) and the superconducting phase($\phi$) as

\[
    r_A= 
\begin{cases}
    e^{-i\phi}X^{-1}\sqrt{cos\alpha \, cos\alpha^{\prime}}& \text{if } |\alpha|<\alpha_c\\
    0,              & \text{if } |\alpha|>\alpha_c
\end{cases}
\]

\begin{equation}
r=X^{-1}(-cos\beta \, sin(\frac{\alpha^{\prime}+\alpha}{2})+isin\beta \, sin(\frac{\alpha^{\prime}-\alpha}{2}))
\end{equation}
where,
\begin{equation}
X=cos\beta \, cos(\frac{\alpha^{\prime}-\alpha}{2})+isin\beta \,cos(\frac{\alpha^{\prime}+\alpha}{2}) 
\end{equation}

\[
   \alpha^{\prime}= 
\begin{cases}
    sin^{-1}(\frac{sin\alpha}{sin\alpha_c})& \text{if } |\alpha|<\alpha_c\\
    sign(\alpha) \, (\frac{\pi}{2}sign(eV_{SD}-E_F)-i \, cosh^{-1}|\frac{sin\alpha}{sin\alpha_c}|),              & \text{if } |\alpha|>\alpha_c
\end{cases}
\]

\[
   \beta= 
\begin{cases}
    cos^{-1}(\frac{eV_{SD}}{\Delta})& \text{if } eV_{SD}<\Delta\\
    -icosh^{-1}(\frac{eV_{SD}}{\Delta}), & \text{if } eV_{SD}>\Delta
\end{cases}
\]

In fig 4(a) $1/G$ x $dI/dV$ is plotted as a function of excitation energy for different equilibrium Fermi energies. It shows different spectrum for low and high Fermi energy. In fig 4(b) $1/G$ x $dI/dV$ is plotted as a function of $V_{BG}$. To compare our experimental result we introduced a finite Gaussian broadening in the system. We assumed the net conductance is effectively the average of $n$ ideal channels each with a uniform doping, the doping amount is equivalent to presence of an extra gate voltage $b_j$.
\begin{equation}
W(b_j)=\frac{1}{\sqrt{2\pi b_{max}}}*e^{-\frac{(b_j-b_D)^2}{2{b_{max}}^2}}
\end{equation}

Where $b_{max}$ is the broadening and $b_D$ is the experimental Dirac point position (voltage where conductance is minimum). In fig 4(c) $1/G$ x $dI/dV$ is plotted for different broadening. It can be noted that normalized value is more than one. However, with Z in the calculation of $1/G$ x $dI/dV$ the normalized differential conductance value will drop down below one. As the effect of barrier is not considered in the calculation of differential conductance, the superconducting state conductance exceeds normal state conductance due to excess current.

\begin{figure*}[ht!]
 \includegraphics[width=18.4cm, height=5cm]{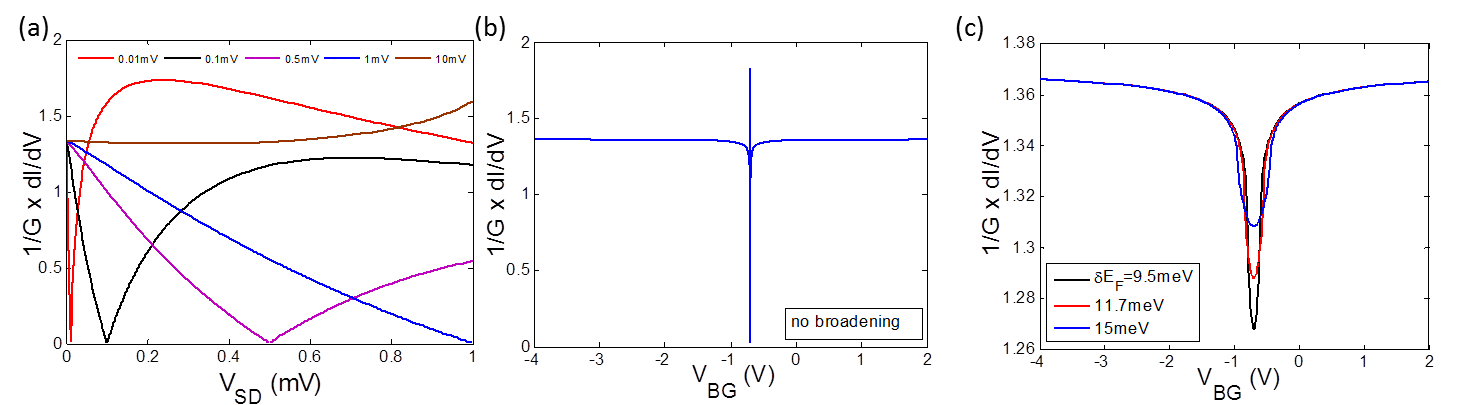}
 \caption{(Color Online) (a) Differential conductance of a NS interface as a function of excitation energy for different Fermi energies. (b) $1/G$ x $dI/dV$ as a function of gate voltage for an ideal GS junction. (c) $1/G$ x $dI/dV$ as a function of gate voltage  for different Fermi energy broadening.}
 \label{fig:example}
\end{figure*}

\bibliography{references}{}
\end{document}